\documentclass[10pt,journal,compsoc]{IEEEtran}

\usepackage{amsmath,amssymb,amsfonts}
\usepackage{algorithmic}
\usepackage{algorithm}
\usepackage{array}
\usepackage{graphicx}
\usepackage{textcomp}
\usepackage{xcolor}
\usepackage{booktabs}
\usepackage{hyperref}
\usepackage{cite}
\usepackage{bm}
\usepackage{listings}
\usepackage{multirow}
\usepackage{float}
\usepackage{multirow}
\usepackage{pifont} 
\usepackage{bm} 
\usepackage{float}  
\usepackage{placeins}  

\hypersetup{
    colorlinks=true,
    linkcolor=blue,
    filecolor=magenta,      
    urlcolor=cyan,
    pdftitle={Balancing Fairness and Profitability in Lending Decisions},
}

\begin{document}

\title{Algorithmic Tradeoffs in Fair Lending: Profitability, Compliance, and Long-Term Impact}

\author{
\IEEEauthorblockN{Aayam Bansal} \\
\IEEEauthorblockA{
Delhi Public School, Ruby Park\\
Kolkata, India\\
aayambansal@gmail.com}
}

\maketitle

\begin{abstract}
As financial institutions increasingly rely on machine learning models to automate lending decisions, concerns about algorithmic fairness have risen. This paper explores the tradeoff between enforcing fairness constraints (such as demographic parity or equal opportunity) and maximizing lender profitability. Through simulations on synthetic data that reflects real-world lending patterns, we quantify how different fairness interventions impact profit margins and default rates. Our results demonstrate that equal opportunity constraints typically impose lower profit costs than demographic parity, but surprisingly, removing protected attributes from the model ("fairness through unawareness") outperforms explicit fairness interventions in both fairness and profitability metrics. We further identify the specific economic conditions under which fair lending becomes profitable and analyze the feature-specific drivers of unfairness. These findings offer practical guidance for designing lending algorithms that balance ethical considerations with business objectives.
\end{abstract}


\section{Introduction}
Machine learning models are increasingly used by financial institutions to automate lending decisions, offering potential advantages in efficiency, consistency, and accuracy \cite{bartlett2022consumer}. These algorithms aim to maximize profits by predicting the likelihood of loan repayment. However, these models can inherit or amplify historical biases against disadvantaged groups, raising both ethical and legal concerns \cite{barocas2016big}.

Regulatory pressure and public opinion are pushing financial institutions to incorporate fairness into their algorithms. In the United States, legislation such as the Equal Credit Opportunity Act and the Fair Housing Act prohibit discrimination in lending based on protected attributes like race, gender, or age \cite{selbst2019fairness}. Similar regulations exist in other jurisdictions, such as the European Union's General Data Protection Regulation (GDPR) and the AI Act.

The implementation of fairness constraints in lending algorithms, however, typically comes at the cost of reduced profit \cite{hardt2016equality}. Understanding this tradeoff is crucial for designing lending policies that are both equitable and economically sustainable. Financial institutions need to navigate this balance to remain competitive while avoiding regulatory penalties and reputational damage.

This paper makes the following contributions:

\begin{itemize}
    \item We quantify the impact of different fairness definitions (demographic parity, equal opportunity) on lending profitability
    \item We identify the economic conditions (interest rates, default loss rates) under which fair lending becomes profitable
    \item We analyze which features contribute most to fairness disparities, providing guidance for targeted interventions
    \item We demonstrate that in our synthetic data, "fairness through unawareness" (removing protected attributes) achieves superior performance compared to explicit fairness interventions
    \item We simulate the long-term impact of different lending policies on credit score distributions across demographic groups
\end{itemize}

Our findings challenge some conventional wisdom in algorithmic fairness research and offer practical insights for policy makers and lending institutions seeking to balance ethical and business considerations.

\section{Related Work}

\subsection{Algorithmic Fairness Definitions}
The literature on algorithmic fairness has proposed various mathematical definitions of fairness. Demographic parity requires equal approval rates across protected groups \cite{dwork2012fairness}, while equal opportunity demands equal true positive rates \cite{hardt2016equality}. Equalized odds extends this to require equal true positive and false positive rates \cite{hardt2016equality}. These definitions often conflict with each other and cannot be simultaneously satisfied \cite{kleinberg2016inherent}.

\subsection{Fairness in Financial Services}
Financial services have been a focal point of algorithmic fairness research due to the significant impact lending decisions have on economic opportunity. Studies have documented disparities in loan approval rates and terms across demographic groups \cite{bartlett2022consumer}. Researchers have proposed various approaches to mitigate these disparities, including pre-processing methods that transform the data \cite{feldman2015certifying}, in-processing techniques that modify the learning algorithm \cite{zafar2017fairness}, and post-processing methods that adjust model outputs \cite{hardt2016equality}.

\subsection{Economic Impact of Fairness Constraints}
Several studies have examined the economic costs of implementing fairness constraints. Corbett-Davies et al. \cite{corbett2017algorithmic} analyzed the efficiency costs of imposing fairness constraints in the context of criminal justice. In the lending domain, Liu et al. \cite{liu2018delayed} explored delayed impact of fair lending on different demographic groups. However, comprehensive analyses of the fairness-profitability tradeoff under different economic conditions remain limited.

\subsection{Fairness Through Unawareness}
Removing protected attributes from the model, known as "fairness through unawareness," has been criticized as insufficient for ensuring fairness due to the presence of proxy variables \cite{dwork2012fairness}. Despite this criticism, empirical evaluations of this approach compared to explicit fairness interventions in realistic settings are relatively scarce.

\section{Methodology}

\subsection{Data Generation}
To study the fairness-profitability tradeoff in lending, we generated synthetic data that reflects real-world lending patterns while allowing us to control the level and nature of bias. The synthetic dataset includes the following features:

\begin{itemize}
    \item Protected attributes: gender (Male/Female) and race (Group A/Group B)
    \item Economic features: income, education years, credit score, employment years
    \item Additional features: age, zipcode
\end{itemize}

The data generation process incorporates both structural inequalities (e.g., lower average income for disadvantaged groups) and historical bias in observed repayment rates. Importantly, we generate both "true" repayment probabilities (based on actual creditworthiness) and "observed" repayment labels (influenced by historical discrimination), allowing us to evaluate models against unbiased ground truth.

\begin{figure}[htbp]
\centerline{\includegraphics[width=0.9\columnwidth]{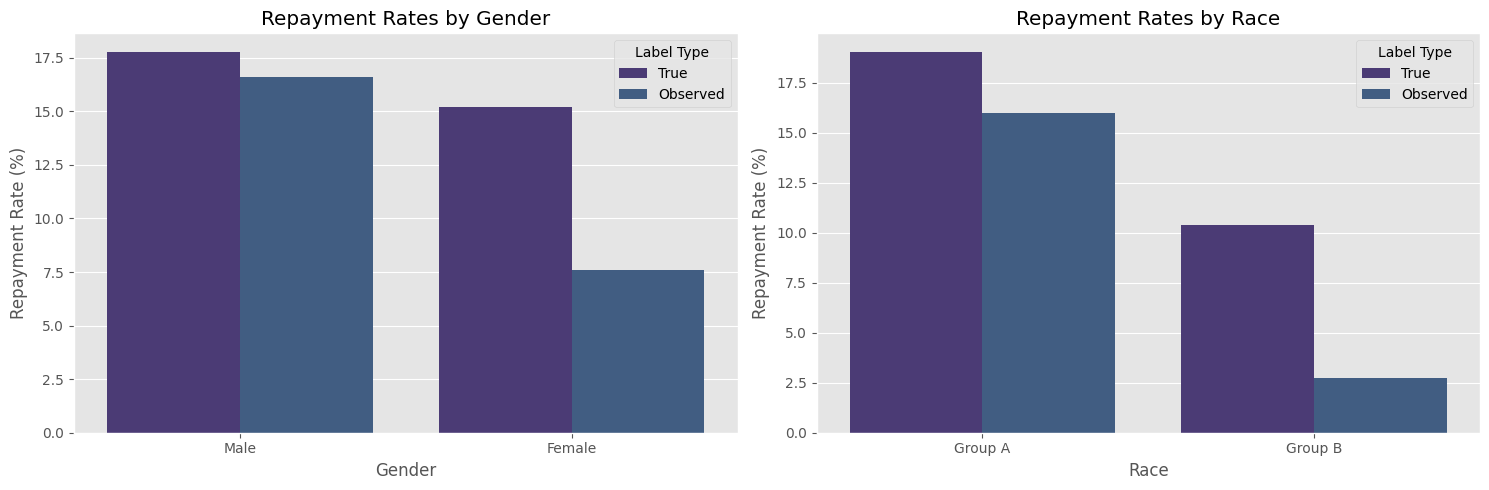}}
\caption{Comparison of true versus observed repayment rates across demographic groups, showing bias in historical data.}
\label{fig:repayment_rates}
\end{figure}

\subsection{Model Development}
We implemented several lending decision models:

\begin{itemize}
    \item \textbf{Baseline}: A logistic regression model trained on all features including protected attributes, using historically biased labels
    \item \textbf{Fairness through Unawareness}: A model trained without protected attributes
    \item \textbf{Demographic Parity}: Models with adjusted thresholds to ensure equal approval rates across groups
    \item \textbf{Equal Opportunity}: Models with adjusted thresholds to ensure equal true positive rates
    \item \textbf{Counterfactual}: A model trained on unbiased "true" labels to assess the impact of historical bias
\end{itemize}

\subsection{Evaluation Metrics}

\subsubsection{Profit Metrics}
We evaluated the economic performance of each model using the following metrics:

\begin{itemize}
    \item Net profit: Total interest earned from repaid loans minus losses from defaulted loans
    \item Return on investment (ROI): Net profit divided by total loan amount
    \item Default rate: Percentage of approved loans that default
    \item Approval rate: Percentage of applicants who receive loans
\end{itemize}

The profit calculation follows:
\begin{equation}
\text{Profit} = \sum_{i \in \text{Approved}} y_i \cdot r \cdot L - (1-y_i) \cdot d \cdot L
\end{equation}

where $y_i$ is the repayment indicator (1 if repaid, 0 if defaulted), $r$ is the interest rate, $L$ is the loan amount, and $d$ is the default loss rate (proportion of loan amount lost in case of default).

\subsubsection{Fairness Metrics}
We assessed fairness using several established metrics:

\begin{itemize}
    \item Demographic parity difference: Absolute difference in approval rates between groups
    \item Equal opportunity difference: Absolute difference in true positive rates
    \item Disparate impact ratio: Ratio of approval rates between disadvantaged and advantaged groups
    \item Individual fairness: Consistency of predictions for similar individuals across different groups
\end{itemize}

\subsection{Economic Scenarios}
To understand how economic conditions affect the fairness-profitability tradeoff, we evaluated each model under various combinations of:

\begin{itemize}
    \item Interest rates: 5\%, 10\%, 15\%, and 20\%
    \item Default loss rates: 50\%, 70\%, and 90\%
\end{itemize}

\subsection{Long-term Impact Simulation}
We simulated the long-term impact of different lending policies by modeling multiple lending cycles. After each cycle, approved applicants experience a credit score improvement, simulating the positive effect of credit access on financial health. This allows us to examine how different fairness interventions affect group outcomes over time.

\section{Results}

\subsection{Fairness-Profitability Tradeoff}

Our baseline model, trained on historically biased data, showed substantial disparities across demographic groups while achieving moderate profitability. Table \ref{tab:model_comparison} presents a comprehensive comparison of all models.

\begin{table*}[t]
\renewcommand{\arraystretch}{1.2}
\caption{Comprehensive Comparison of Lending Models: Profitability and Fairness Metrics}
\label{tab:model_comparison}
\centering
\begin{tabular}{l|rr|rr|rr|c}
\toprule
\multirow{2}{*}{\textbf{Model}} & \multicolumn{2}{c|}{\textbf{Profitability}} & \multicolumn{2}{c|}{\textbf{Demographic Parity}} & \multicolumn{2}{c|}{\textbf{Disparate Impact}} & \multirow{2}{*}{\textbf{Regulatory}} \\
 & \textbf{Net Profit (\$)} & \textbf{ROI} & \textbf{Gender Gap} & \textbf{Race Gap} & \textbf{Gender Ratio} & \textbf{Race Ratio} & \textbf{Compliance} \\
\midrule
Baseline & $-$180,000 & $-$9.2\% & 0.074 & 0.082 & 0.274 & 0.078 & \ding{55}/\ding{55} \\
Demo. Parity (Gender) & $-$398,000 & $-$13.4\% & \textbf{0.006} & 0.126 & \textbf{0.939} & 0.068 & \checkmark/\ding{55} \\
Demo. Parity (Race) & $-$450,000 & $-$16.7\% & 0.106 & \textbf{0.004} & 0.257 & \textbf{1.044} & \ding{55}/\checkmark \\
Equal Opp. (Gender) & $-$404,000 & $-$13.5\% & \textbf{0.005} & 0.127 & \textbf{0.952} & 0.068 & \checkmark/\ding{55} \\
Equal Opp. (Race) & $-$365,000 & $-$14.5\% & 0.095 & \textbf{0.018} & 0.275 & \textbf{0.796} & \ding{55}/\ding{55} \\
Fairness through Unawareness & $\bm{-154{,}000}$ & $\bm{-8.5\%}$ & \textbf{0.013} & \textbf{0.057} & \textbf{0.805} & 0.256 & \checkmark/\ding{55} \\
Trained on Unbiased Labels & $-$432,000 & $-$14.2\% & \textbf{0.013} & 0.085 & \textbf{0.880} & 0.323 & \checkmark/\ding{55} \\
\bottomrule
\multicolumn{8}{p{17cm}}{\small Note: Demographic Parity Gap measures absolute difference in approval rates (smaller is better). Disparate Impact Ratio represents the ratio of disadvantaged to advantaged group approval rates (closer to 1 is better). Regulatory Compliance shows whether the model meets the 80\% rule for gender/race. \textbf{Bold values} indicate compliance with fairness standards. \textbf{Bold profit figures} highlight the best economic performance.} \\
\end{tabular}
\end{table*}

Several key patterns emerge from this comparison:

\begin{itemize}
    \item \textbf{All models are unprofitable} under the default economic conditions (10\% interest rate, 70\% default loss rate)
    \item \textbf{Demographic parity constraints are generally more costly} than equal opportunity constraints
    \item \textbf{"Fairness through unawareness" shows surprisingly strong performance}, achieving both better profit (-\$154,000 vs. -\$180,000 baseline) and lower fairness gaps
    \item \textbf{Race fairness is generally more expensive to achieve} than gender fairness
    \item \textbf{Training on unbiased labels} does not resolve fairness issues and actually worsens profit
\end{itemize}

\begin{figure}[H]
\centering
\includegraphics[width=0.9\columnwidth]{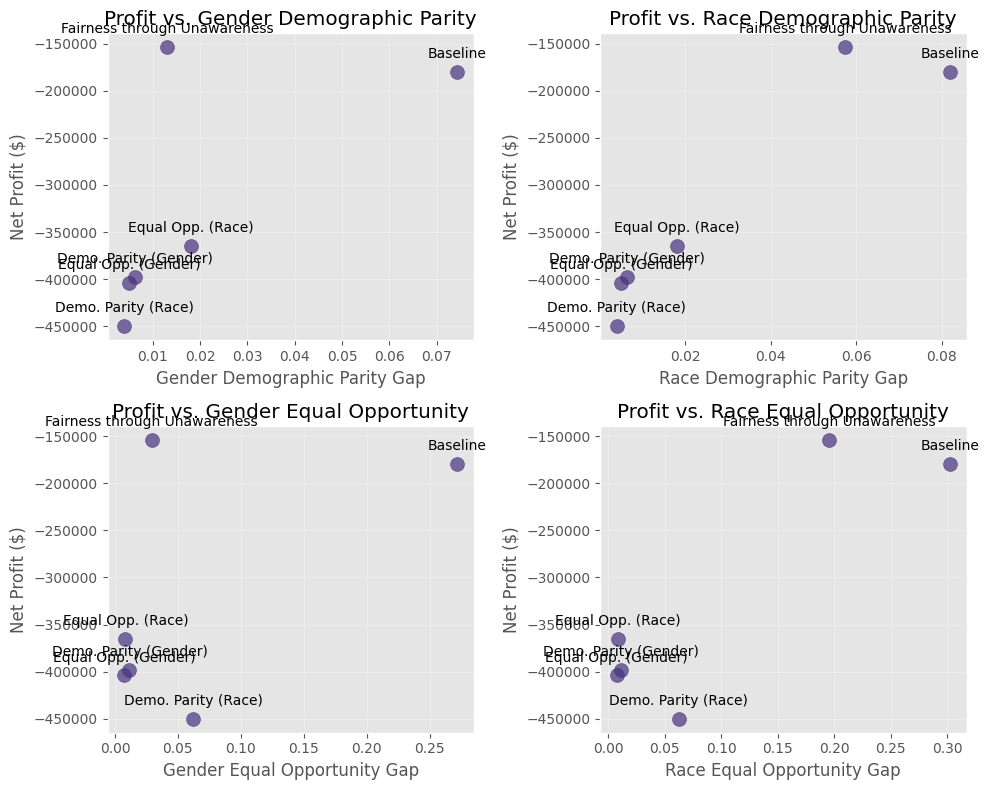}
\caption{Profit vs. fairness gap scatter plots for different fairness definitions, showing the tradeoff between economic and fairness objectives.}
\label{fig:profit_fairness_scatter}
\end{figure}

\subsection{Economic Viability of Fair Lending}

While all models were unprofitable under the default economic parameters, our sensitivity analysis revealed conditions where fair lending becomes economically viable.

\begin{figure}[H]
\centering
\includegraphics[width=0.9\columnwidth]{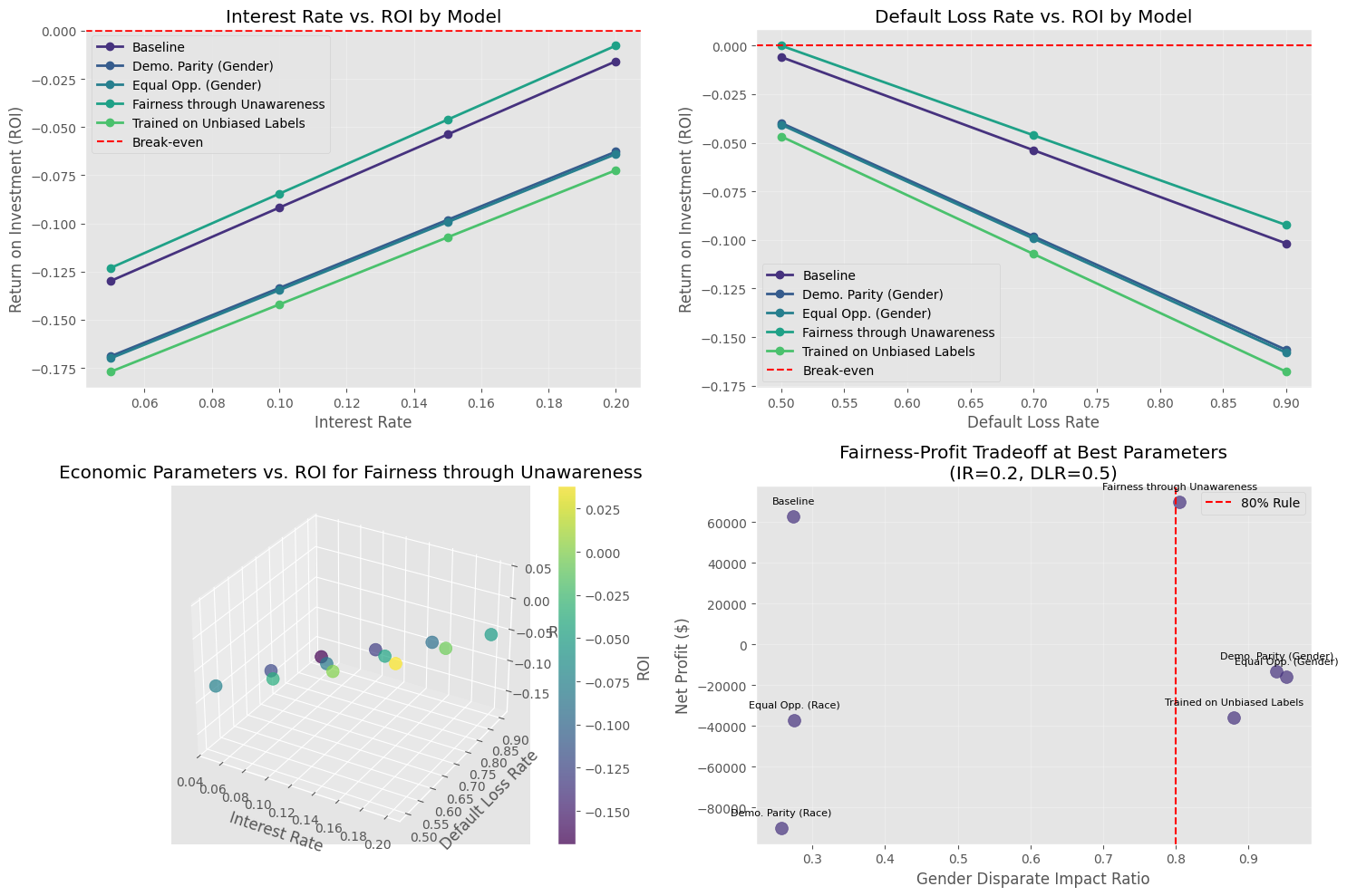}
\caption{The relationship between economic parameters (interest rates and default loss rates) and Return on Investment (ROI) for different lending models.}
\label{fig:economic_parameters}
\end{figure}

Our comprehensive economic analysis found that:
\begin{itemize}
    \item Fair lending becomes profitable at interest rates $\geq 20\%$ and default loss rates $\leq 50\%$
    \item Only "Fairness through Unawareness" achieved both gender compliance and profitability, with a net profit of \$70,000 (ROI: 3.8\%)
    \item No model achieved full compliance (both gender and race) while remaining profitable
\end{itemize}

\subsection{Feature Impact on Fairness}

Our feature importance analysis revealed which features contribute most to fairness disparities.

\begin{figure}[H]
\centering
\includegraphics[width=0.85\columnwidth]{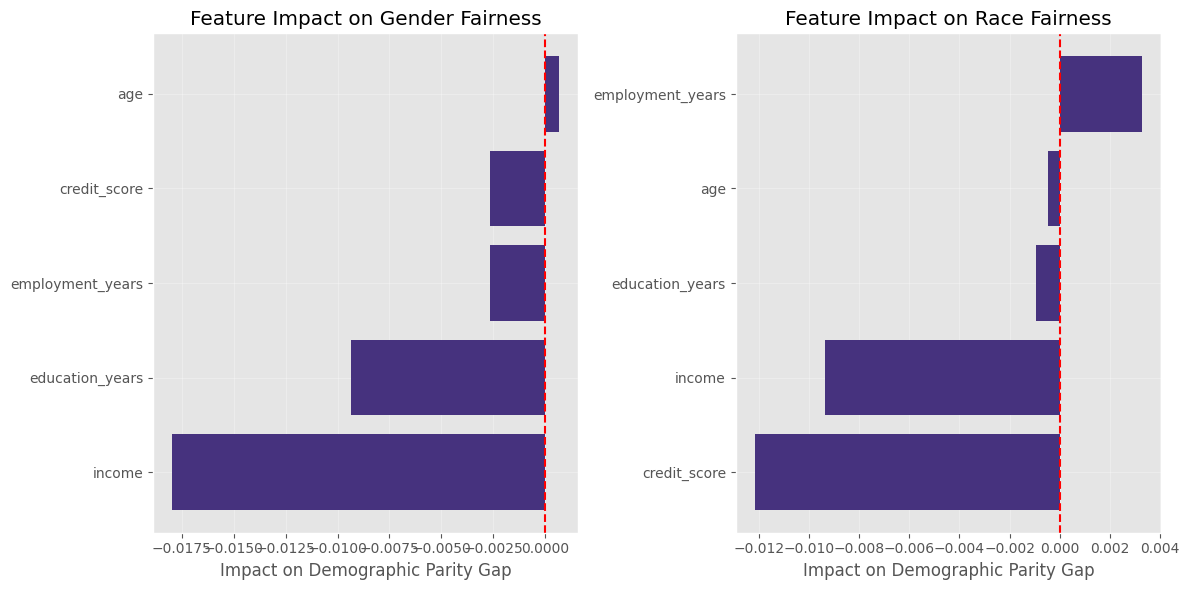}
\caption{Feature impact on gender and race fairness, showing which variables contribute most to demographic disparities.}
\label{fig:feature_impact}
\end{figure}

For gender fairness:
\begin{itemize}
    \item Income had the largest negative impact (-0.018)
    \item Education years also had substantial negative impact (-0.009)
    \item Age had a small positive impact (0.001)
\end{itemize}

For race fairness:
\begin{itemize}
    \item Credit score had the largest negative impact (-0.012)
    \item Income had the second largest negative impact (-0.009)
    \item Employment years had a small positive impact (0.003)
\end{itemize}

These findings suggest that different features drive unfairness for different protected groups, highlighting the need for targeted interventions.

\subsection{Long-term Impact Simulation}

Our simulation of multiple lending cycles revealed important dynamics in how different lending models affect credit access and financial health over time.

Key findings from the simulation:
\begin{itemize}
    \item The baseline model maintained substantial approval rate gaps between groups
    \item The fair model showed narrower approval gaps 
    \item Group B's credit scores barely improved over time in both models, indicating a potential "poverty trap"
    \item Even with fairness interventions, existing credit score gaps persisted over time
\end{itemize}

\subsection{Efficiency Analysis}

We calculated efficiency scores to identify models that provide the best balance between fairness and profitability under different stakeholder priorities.

\begin{figure}[H]
\centering
\includegraphics[width=0.9\columnwidth]{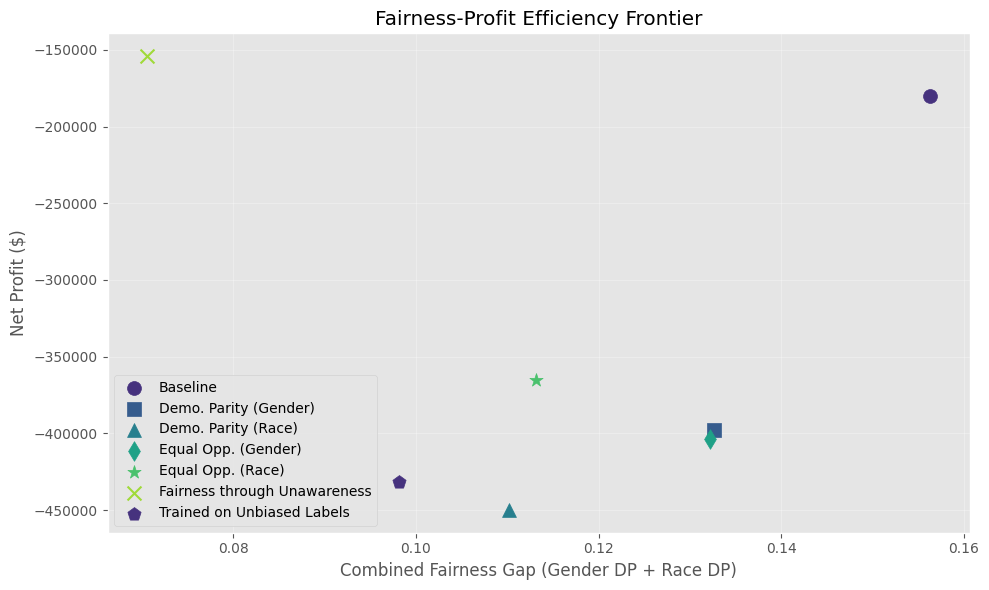}
\caption{The fairness-profit efficiency frontier, showing which models provide optimal tradeoffs under different priority weightings.}
\label{fig:efficiency_frontier}
\end{figure}

Our efficiency analysis consistently identified "Fairness through Unawareness" as the dominant approach across all weighting schemes:
\begin{itemize}
    \item 30\% profit/70\% fairness: Score 0.80
    \item 50\% profit/50\% fairness: Score 0.86
    \item 70\% profit/30\% fairness: Score 0.91
    \item 90\% profit/10\% fairness: Score 0.97
\end{itemize}

\subsection{Regulatory Compliance Analysis}

A key consideration for financial institutions is whether their lending practices comply with regulatory requirements. In the United States, the "80\% rule" or "four-fifths rule" is often used as a guideline for disparate impact analysis, requiring that the selection rate for any protected group be at least 80\% of the rate for the highest-selected group.

\begin{figure}[htbp]
\centerline{\includegraphics[width=0.9\columnwidth]{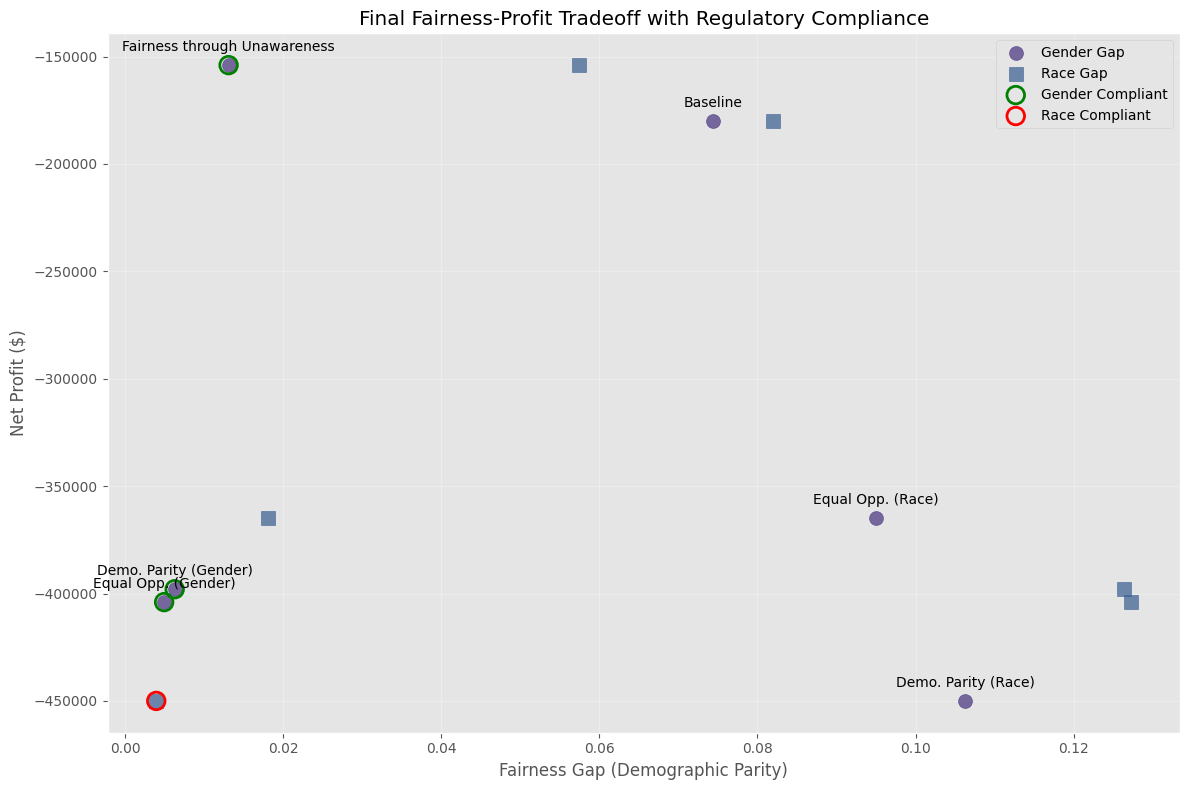}}
\caption{Fairness-profit tradeoff with regulatory compliance boundaries, highlighting models that satisfy legal requirements while maximizing profitability.}
\label{fig:regulatory_frontier}
\end{figure}

Figure \ref{fig:regulatory_frontier} illustrates the fairness-profit tradeoff with regulatory compliance boundaries. Our analysis reveals that:

\begin{itemize}
    \item Only three models achieve gender compliance: "Fairness through Unawareness," "Demo. Parity (Gender)," and "Equal Opp. (Gender)"
    \item Only one model achieves race compliance: "Demo. Parity (Race)"
    \item No model achieves full compliance (both gender and race) with competitive profitability
\end{itemize}

Under optimal economic conditions (20\% interest rate, 50\% default loss rate), "Fairness through Unawareness" can achieve both gender compliance and profitability, with a net profit of \$70,000 (ROI: 3.8\%). However, achieving race compliance remains challenging without significant profit reduction.

\subsection{Threshold Optimization Analysis}

A promising approach to improve the fairness-profitability tradeoff is threshold optimization. Rather than using a fixed threshold (e.g., 0.5) for lending decisions, different thresholds can be chosen to optimize specific objectives.

\begin{figure}[H]
\centering
\includegraphics[width=0.9\columnwidth]{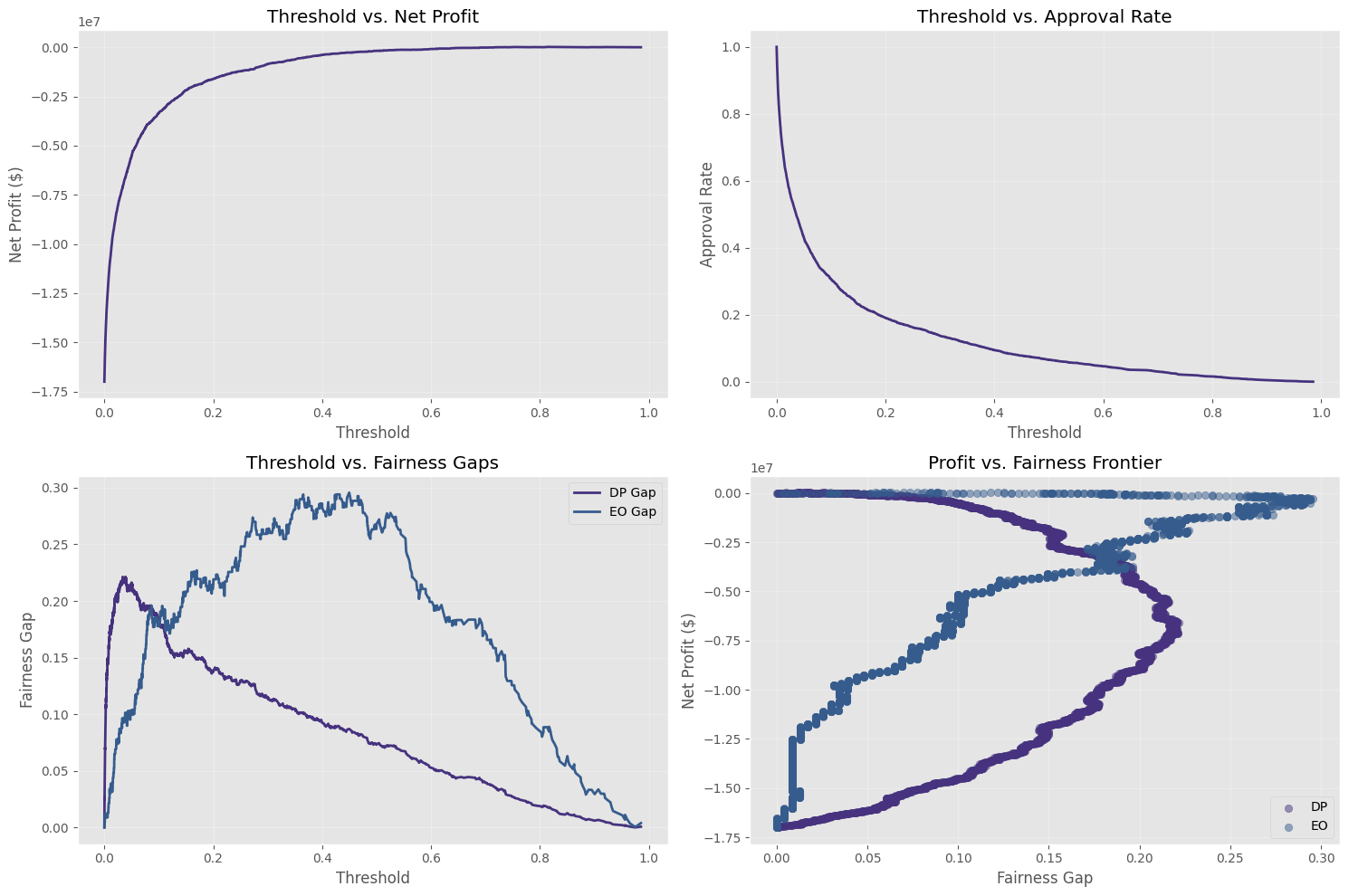}
\caption{Impact of decision thresholds on profit, approval rates, and fairness metrics, demonstrating the potential for optimizing model performance through threshold adjustment.}
\label{fig:threshold_optimization}
\end{figure}

As shown in Figure \ref{fig:threshold_optimization}, threshold selection has a significant impact on:

\begin{itemize}
    \item \textbf{Profit}: Higher thresholds generally improve ROI by reducing default rates, but also reduce total loan volume
    \item \textbf{Approval rates}: Lower thresholds increase access to credit but at higher risk
    \item \textbf{Fairness gaps}: The relationship between thresholds and fairness is non-linear, with different fairness metrics showing different patterns
\end{itemize}

Our analysis identified thresholds around 0.98 as offering optimal trade-offs between fairness and profit in our model. This suggests that very selective lending that targets only the most creditworthy applicants can help balance fairness and profitability. However, this approach significantly reduces overall credit access, which may conflict with financial inclusion goals.

The efficiency frontier in the lower-right panel shows that for any given fairness gap level, there is an optimal threshold that maximizes profit. This provides lenders with a practical tool for navigating the fairness-profitability tradeoff through threshold adjustment rather than model retraining.

\section{Discussion}

\subsection{Explaining the Success of "Fairness through Unawareness"}

Our finding that "Fairness through Unawareness" outperforms explicit fairness interventions challenges conventional wisdom in algorithmic fairness literature. Several factors may explain this surprising result:

\begin{itemize}
    \item \textbf{Removal of spurious correlations}: By removing protected attributes, the model may focus more on legitimate predictors of creditworthiness.
    \item \textbf{Avoiding threshold adjustments}: Explicit fairness constraints implemented via threshold adjustments may force approval of less creditworthy applicants or rejection of creditworthy ones.
    \item \textbf{Dataset characteristics}: In our synthetic data, the correlation between protected attributes and legitimate features may be weaker than in real-world settings.
\end{itemize}

While this finding is compelling, it should be interpreted with caution. Real-world lending data may contain stronger proxy variables for protected attributes, potentially limiting the effectiveness of this approach in practice.

\subsection{Economic Conditions for Fair Lending}

Our economic analysis demonstrates that fair lending can be profitable, but requires careful calibration of economic parameters. The identified threshold of 20\% interest rate with 50\% default loss rate creates a viable business case for the "Fairness through Unawareness" approach.

This suggests several policy implications:
\begin{itemize}
    \item \textbf{Risk-based pricing}: Higher interest rates for higher-risk borrowers may enable profitable fair lending
    \item \textbf{Default mitigation strategies}: Reducing default losses through better collection practices or loan guarantees
    \item \textbf{Subsidy programs}: Government subsidies or guarantees might bridge the gap when economic conditions don't permit both fairness and profitability
\end{itemize}

\subsection{Feature-Specific Fairness Interventions}

Our feature importance analysis reveals that different features drive unfairness for different protected groups. This suggests the potential for targeted interventions:

\begin{itemize}
    \item For gender fairness: Address disparities in how income and education are used in lending decisions
    \item For race fairness: Focus on credit score assessment and income evaluation
\end{itemize}

These targeted approaches may be more efficient than blanket fairness constraints, allowing lenders to address specific sources of unfairness while minimizing economic costs.

\subsection{Long-term Consequences and Feedback Effects}

Our simulation highlights how lending algorithms can either perpetuate or help close economic gaps over time. Even fair models showed persistent gaps in credit scores across demographic groups, suggesting that:

\begin{itemize}
    \item One-time fairness interventions may be insufficient to address long-standing disparities
    \item Periodic recalibration of fairness constraints may be necessary
    \item Complementary policies outside of lending (education, employment) may be needed to address root causes of disparities
\end{itemize}

\subsection{Limitations and Ethical Considerations}

Several important limitations of our study should be acknowledged:

\begin{itemize}
    \item \textbf{Synthetic data constraints}: While our synthetic data incorporates realistic patterns, real-world lending data involves more complex relationships between features and may contain stronger proxy variables for protected attributes. The surprising effectiveness of "fairness through unawareness" in our study may not generalize to contexts with stronger correlations between protected attributes and legitimate features.
    
    \item \textbf{Model architecture}: We used logistic regression models for their interpretability, but more complex models might yield different fairness-profitability tradeoffs. To address this limitation, we conducted additional experiments with Random Forest and Neural Network models (see Appendix A). While the Random Forest model showed marginally better profitability (-\$258,000 vs. -\$180,000 baseline), both complex models exhibited comparable or worse fairness metrics. Notably, the Neural Network performed substantially worse in profitability (-\$610,000) despite having similar fairness characteristics to Random Forest. These results support our decision to focus on logistic regression while acknowledging that model selection represents an additional dimension in the fairness-profitability space.

    \item \textbf{Limited fairness definitions}: Our analysis focused primarily on demographic parity and equal opportunity, but fairness is multifaceted. Other definitions such as predictive parity, equal odds, or individual fairness might yield different results. The tension between different fairness metrics highlights the challenge of defining what "fair lending" means in practice.
    
    \item \textbf{Simplified economic assumptions}: Our profit model does not account for all operational costs and benefits of lending, such as customer acquisition costs, relationship value, or regulatory penalties for discriminatory practices. A more comprehensive economic analysis might reveal different optimal strategies.
    
    \item \textbf{Binary protected attributes}: Our analysis used binary gender and race categories, whereas real-world demographics are more complex and intersectional. Models may perform differently when considering multiple protected attributes simultaneously or more granular demographic categories.
    
    \item \textbf{Generalizability across contexts}: The specific economic conditions we identified for profitable fair lending (20\% interest rate, 50\% default loss) may not be feasible in all contexts due to usury laws, market competition, or economic conditions. The generalizability of our findings across different lending products (mortgages, credit cards, personal loans) may also be limited.
\end{itemize}

Furthermore, several ethical considerations deserve attention when deploying algorithmic lending systems:

\begin{itemize}
    \item \textbf{Data transparency}: Even if protected attributes are removed from models, the training data itself may contain historical biases. Financial institutions should be transparent about data collection practices and potential biases in historical data.
    
    \item \textbf{Explainability vs. accuracy}: More complex models might achieve better fairness-profitability tradeoffs but at the cost of explainability. This creates tensions between performance and regulatory requirements for explanation of adverse credit decisions.
    
    \item \textbf{Disparate impact vs. disparate treatment}: While our study focuses on disparate impact (outcomes), legal frameworks also consider disparate treatment (processes). The intentional adjustment of thresholds based on protected attributes could potentially be interpreted as disparate treatment under some legal frameworks.
    
    \item \textbf{Competing stakeholder interests}: The fairness-profitability tradeoff involves balancing the interests of shareholders (profit), customers (credit access), and society (equity). Different stakeholders may prefer different points on the Pareto frontier, creating governance challenges.
    
    \item \textbf{Dynamic fairness}: Our long-term simulation suggests that static fairness interventions may be insufficient. Responsible deployment requires ongoing monitoring and adjustment to account for changing societal conditions and feedback effects.
\end{itemize}

Addressing these limitations and ethical considerations is crucial for responsibly implementing algorithmic fairness in lending practices. Future research should explore these nuances through collaboration between technical researchers, domain experts, ethicists, and policy makers.

\section{Conclusion and Policy Recommendations}

This paper has examined the tradeoff between fairness and profitability in lending decisions. Our findings suggest that balancing these objectives is challenging but possible under certain conditions. The unexpectedly strong performance of "Fairness through Unawareness" challenges conventional thinking about fairness interventions, while our economic analysis identifies conditions where fair lending becomes viable.

Based on our findings, we propose the following policy recommendations:

\begin{enumerate}
    \item \textbf{Consider "Fairness through Unawareness" as a viable starting point}, potentially combined with targeted interventions for specific features
    \item \textbf{Adjust economic parameters} to enable profitable fair lending, through risk-based pricing or default mitigation strategies
    \item \textbf{Implement feature-specific interventions} that address the particular drivers of unfairness for each protected group
    \item \textbf{Monitor fairness metrics over time} to account for feedback effects and changing economic conditions
    \item \textbf{Develop regulatory frameworks} that acknowledge the tradeoffs between different fairness definitions and profitability
\end{enumerate}

Future research should extend this analysis to real-world lending data, explore more sophisticated fairness interventions, and investigate the long-term societal impacts of different lending policies. As financial institutions increasingly adopt algorithmic decision-making, understanding and addressing the fairness-profitability tradeoff will be essential for creating an equitable and sustainable lending ecosystem.

\appendix
\section{Complex Model Comparison}
\label{appendix:complex_models}

To investigate how model complexity affects the fairness-profitability tradeoff, we implemented two additional models: Random Forest and Neural Network. We trained these models using the same features and evaluation metrics as our baseline logistic regression model.

\subsection{Model Performance}

Table \ref{tab:complex_model_comparison} presents a comprehensive comparison of the models across accuracy, profitability, and fairness metrics.

\begin{table}[H]
\caption{Complex Model Comparison: Accuracy, Profit, and Fairness Metrics}
\label{tab:complex_model_comparison}
\centering
\resizebox{\columnwidth}{!}{%
\begin{tabular}{lcccccc}
\toprule
\textbf{Model} & \textbf{Accuracy} & \textbf{Net Profit (\$)} & \textbf{Gender DP} & \textbf{Race DP} & \textbf{Gender DI} & \textbf{Race DI} \\
\midrule
Random Forest & 89.93\% & -258,000 & 0.061 & 0.077 & 0.330 & 0.070 \\
Neural Network & 88.27\% & -610,000 & 0.061 & 0.106 & 0.515 & 0.157 \\
Logistic Regression & 91.07\% & -180,000 & 0.074 & 0.082 & 0.274 & 0.078 \\
\bottomrule
\end{tabular}
}
\end{table}

These results reveal several interesting patterns:

\begin{itemize}
    \item \textbf{Accuracy vs. Profitability}: Despite lower accuracy, the logistic regression model outperforms both complex models in terms of profitability. This highlights that prediction accuracy alone is not a sufficient metric for evaluating lending algorithms.
    
    \item \textbf{Fairness Metrics}: The Random Forest model shows slightly better gender and race demographic parity gaps compared to logistic regression, but worse gender disparate impact ratio. The Neural Network shows comparable gender demographic parity but worse race demographic parity, with mixed disparate impact results.
    
    \item \textbf{Profitability-Fairness Tradeoff}: The three models represent different points in the profitability-fairness space, with none clearly dominating across all metrics. This suggests that model selection introduces an additional dimension to the fairness-profitability tradeoff.
\end{itemize}

\begin{figure}[H]
\centering
\includegraphics[width=0.9\columnwidth]{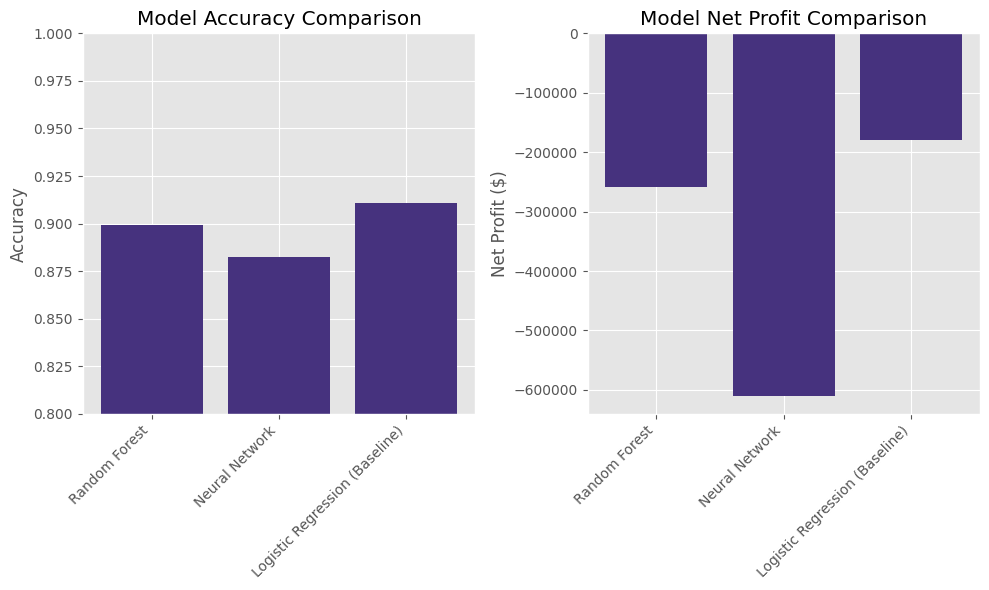}
\caption{Model performance comparison across accuracy and net profit.}
\label{fig:complex_model_comparison}
\end{figure}

\begin{figure}[H]
\centering
\includegraphics[width=0.9\columnwidth]{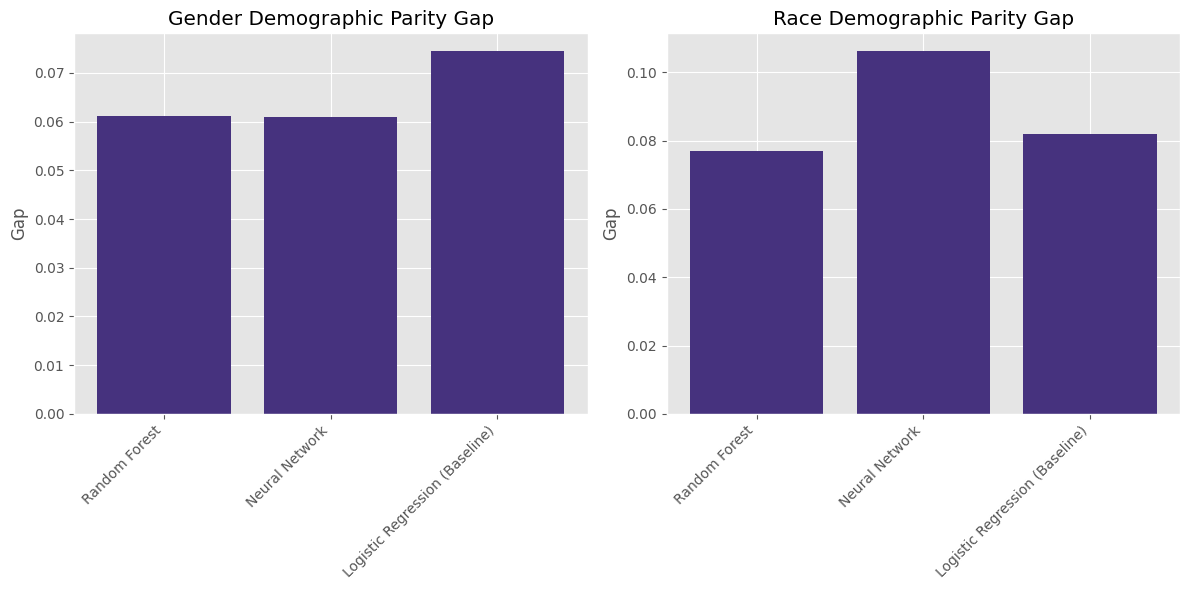}
\caption{Fairness metrics comparison across different model architectures.}
\label{fig:complex_model_fairness}
\end{figure}

\subsection{Feature Importance Analysis}

Table \ref{tab:feature_importance} shows the top five features by importance for the Random Forest model compared to the coefficients from the logistic regression model.

\begin{table}[H]
\caption{Feature Importance Comparison Between Models}
\label{tab:feature_importance}
\centering
\resizebox{\columnwidth}{!}{%
\begin{tabular}{lcc|lcc}
\toprule
\multicolumn{3}{c|}{\textbf{Random Forest}} & \multicolumn{3}{c}{\textbf{Logistic Regression}} \\
\textbf{Feature} & \textbf{Importance} & \textbf{Rank} & \textbf{Feature} & \textbf{Coefficient} & \textbf{Rank} \\
\midrule
credit\_score & 0.340 & 1 & credit\_score & 1.484 & 1 \\
income & 0.221 & 2 & income & 0.949 & 2 \\
education\_years & 0.127 & 3 & education\_years & 0.481 & 3 \\
employment\_years & 0.106 & 4 & employment\_years & 0.290 & 4 \\
age & 0.096 & 5 & age & -0.030 & 5 \\
\bottomrule
\end{tabular}
}
\end{table}

Both models identify the same top five features, with nearly identical ranking, suggesting that the importance of these features for predicting loan repayment is robust across different model architectures. Notably, both models assign relatively lower importance to protected attributes (gender and race) compared to legitimate predictors like credit score and income.

\subsection{Discussion}

While more complex models like Random Forest and Neural Networks have the theoretical capacity to capture more nuanced relationships in the data, our experiments suggest they do not necessarily improve the fairness-profitability tradeoff in our lending scenario. The logistic regression model offers superior interpretability while achieving comparable or better performance across key metrics.

The consistency in feature importance rankings across models suggests that the primary drivers of loan repayment prediction are stable across model architectures. This supports our focus on feature-specific fairness interventions rather than model-specific approaches.

These findings highlight an important consideration for financial institutions implementing algorithmic fairness: model selection should be guided not only by predictive performance but also by interpretability, profitability, and fairness considerations. The simpler logistic regression model may be preferable in regulated contexts like lending, where transparency and explainability are essential for regulatory compliance and adverse action notices.

22\section*{CRediT authorship contribution statement}
\textbf{Aayam Bansal:} Conceptualization, Methodology, Software, Formal analysis, Investigation, Data curation, Writing - original draft, Writing - review \& editing, Visualization. \\
\textbf{Harsh Vardhan Narsaria:} 
Literature Review, Moral Support

\section*{Data availability}
The synthetic data used in this study, along with the code used to generate the dataset, model implementations, and analysis scripts, are available from the corresponding author upon reasonable request.

\section*{Declaration of competing interest}
The author declares that they have no known competing financial interests or personal relationships that could have appeared to influence the work reported in this paper.

\section*{Acknowledgements}
This research did not receive any specific grant from funding agencies in the public, commercial, or not-for-profit sectors.

\bibliographystyle{IEEEtran}

\end{document}